\documentclass[12pt]{article}

\usepackage{amsmath, amsthm, amssymb}
\usepackage{geometry}
\newtheorem{theorem}{Theorem}
\newtheorem{lemma}[theorem]{Lemma}
\newtheorem{proposition}[theorem]{Proposition}
\newtheorem{corollary}[theorem]{Corollary}

\newtheorem{remark}[theorem]{Remark}

\newtheorem{example}[theorem]{Example}

\title{One-Sided Transfer and Absorbing Structure for Markov Kernels}

\author{
Jean-Gabriel~Attali\thanks{Contact: jean-gabriel.attali@devinci.fr} \\
\small De Vinci Higher Education, De Vinci Research Center, Paris, France
}

\date{}

\begin{document} 
\maketitle

\begin{abstract}
For two $\sigma$-finite invariant measures of a Markov kernel, we construct a nonnegative transfer measure that carries mass from the singular part of their relative Lebesgue decomposition into the measure class of the reference measure. When the reference measure is conservative, this transfer is zero. When the singular component is nonzero, vanishing is equivalent to separate invariance of the two components and their placement on disjoint absorbing sets. Under a conservative $\sigma$-finite invariant measure, indecomposability and irreducibility with respect to that measure coincide. In continuous time, the transfers form a positive cocycle, which vanishes whenever a normalized resolvent is conservative.

\end{abstract}

\section{Introduction}

There are two natural notions of absence of decomposition for a Markov
chain. In the terminology of Breiman~\cite{Breiman}, a Markov kernel is
\emph{indecomposable} if there are no two disjoint nonempty absorbing
measurable sets. Irreducibility with respect to a reference measure
requires every set of positive measure to be accessible from every
starting point; see Nummelin~\cite{Nummelin}, Revuz~\cite{Revuz}, and
Meyn and Tweedie~\cite{MeynTweedie}. Irreducibility implies
indecomposability, but the converse fails in general.

For finite invariant measures, mass balance often removes this
distinction. In the sigma-finite setting, conservativity plays a similar
role.

Let $\mu$ and $\nu$ be sigma-finite invariant measures and write
\[
\nu=h\mu+\nu_s,
\qquad
\nu_s\perp\mu.
\]
The two components need not be separately invariant. Their evolution is nevertheless
one-sided: there exists a nonnegative measure $\alpha\ll\mu$ such that
\[
\nu_sP=\nu_s+\alpha,
\qquad
(h\mu)P+\alpha=h\mu.
\]
Thus the singular component may acquire mass in the measure class of $\mu$, whereas
the absolutely continuous component, together with the transferred mass, reconstructs
its initial mass.

We give a spatially homogeneous nearest-neighbor random walk on a
two-layer strip with a nonconservative invariant reference measure and
nonzero transfer. By contrast, if $\mu$ is conservative, then
$\alpha=0$, even when both components have infinite mass.
Conservative Markov kernels with sigma-finite invariant measures have
long been studied in ergodic and Markov operator theory; see
Foguel~\cite{Foguel1969}, Horowitz~\cite{Horowitz1968}, and Lin~\cite{Lin1970}.
Hence conservativity preserves the relative Lebesgue decomposition of invariant measures.

An analogous statement holds for absorbing sets. If $A$ is absorbing,
the splitting $\mu=1_A\mu+1_{A^c}\mu$ has an associated transfer given
by the mass entering $A$ from $A^c$. Its vanishing is equivalent to
$P1_A=1_A$ $\mu$-a.e. We call this property, when it holds for every
absorbing set, \emph{absorbing harmonicity}. Conservativity implies
absorbing harmonicity.

For a measurable set $B$, the zero-potential set
$Z_B=\{x:R_\theta(x,B)=0\}$ is absorbing. Combining this observation
with absorbing harmonicity yields, for every nonzero sigma-finite
invariant measure $\mu$ with respect to which $P$ is conservative,
\[
        P\text{ indecomposable}
        \quad\Longleftrightarrow\quad
        P\text{ is $\mu$-irreducible}.
\]
Without conservativity the converse implication can fail: we give an
indecomposable kernel admitting no irreducibility measure. The result
also covers conservative infinite-measure regimes, including null
Harris recurrence.

If
$\nu_s\neq0$, zero transfer is equivalent to separate invariance of
$h\mu$ and $\nu_s$, and to the existence of disjoint absorbing carriers
for the reference and singular components. Iterating the transfer gives
a potential--invariant decomposition of the excessive measure $h\mu$,
with defect $\alpha$. This is related to the classical Riesz theory of
excessive measures; see Getoor and Glover~\cite{GetoorGlover1984},
Fitzsimmons and Getoor~\cite{FitzsimmonsGetoor1992}, and
Getoor~\cite{Getoor1990}. 
The present purpose is the identification of its
defect as the one-sided transfer generated by the singular component
of another invariant measure, together with the conservative vanishing
and absorbing-separation consequences.

Mutually singular sigma-finite invariant measures admit disjoint
absorbing carriers. For invariant probabilities, the Jordan
decomposition of the difference of two distinct invariant measures
reduces nonuniqueness to such a singular pair. Consequently, an
indecomposable kernel admits at most one invariant probability measure.

Chen and Tweedie~\cite{ChenTweedie} treat orthogonality and absorbing sets in similar settings. A complementary topological
viewpoint, based on closed absorbing sets and the supports of invariant
measures, is developed in Attali~\cite{AttaliClosedAbsorbing}.

In continuous time, the transfer family satisfies
$\alpha_{s+t}=\alpha_sP_t+\alpha_t$. Normalized resolvents encode
accessibility and absorbing geometry. Conservativity of a normalized
resolvent leads to $\alpha_t=0$ for every $t\ge0$, and zero transfer is
again equivalent to separate invariance and common absorbing
separation.

Section~2 introduces
the one-sided transfer associated with the relative Lebesgue decomposition
of two invariant measures and identifies two sufficient conditions for it to vanish: finiteness and, more generally, conservativity. Sections~3--4
develop the absorbing-set viewpoint through transfer across absorbing
boundaries, zero-potential sets, irreducibility, and absorbing separation.
Section~5 characterizes zero transfer in terms of separate invariance and absorbing separation. Section~6 carries the same structure over to continuous time.

\section{One-sided transfer between invariant measures}
\label{sec:transfer}

Let $P$ be a Markov kernel on a measurable space $(E,\mathcal E)$.
Throughout this section, $\mu$ and $\nu$ are sigma-finite invariant
measures for $P$, with Lebesgue decomposition
\[
        \nu=h\mu+\nu_s,
        \qquad
        \nu_s\perp\mu.
\]
We first identify the associated one-sided transfer and then give two
conditions forcing it to vanish: finiteness of either component and,
more generally, conservativity of the reference measure $\mu$.

\subsection{One-sided transfer}

The relative Lebesgue decomposition of two invariant measures has a
necessarily one-sided evolution.

\begin{theorem}[One-sided transfer]
\label{thm:transfer}
Let $\mu$ and $\nu$ be sigma-finite invariant measures and write
\[
\nu=h\mu+\nu_s,
\qquad
\nu_s\perp\mu.
\]
Then there exists a unique nonnegative measure $\alpha\ll\mu$ such that
\[
\nu_sP=\nu_s+\alpha,
\qquad
(h\mu)P+\alpha=h\mu.
\]
\end{theorem}

\begin{proof}
Since $\mu P=\mu$, the kernel preserves absolute continuity with respect
to $\mu$. Indeed, if $\mu(A)=0$, then
\[
0=\mu P(A)=\int_E P(x,A)\,\mu(dx),
\]
so $P(x,A)=0$ for $\mu$-almost every $x$. Thus
$\lambda P\ll\mu$ whenever $\lambda\ll\mu$, and in particular
\[
(h\mu)P\ll\mu.
\]

Using invariance of $\nu$,
\[
\nu=(h\mu)P+\nu_sP.
\]
Moreover, $\nu_sP\leq\nu$, hence $\nu_sP$ is sigma-finite. Write its
Lebesgue decomposition relative to $\mu$ as
\[
\nu_sP=g\mu+\rho,
\qquad
\rho\perp\mu.
\]
Then
\[
\nu=(h\mu)P+g\mu+\rho,
\]
with $(h\mu)P+g\mu\ll\mu$ and $\rho\perp\mu$. Uniqueness of the
Lebesgue decomposition of
\[
\nu=h\mu+\nu_s
\]
therefore gives
\[
\rho=\nu_s,
\qquad
(h\mu)P+g\mu=h\mu.
\]
Setting $\alpha:=g\mu$ yields
\[
\nu_sP=\nu_s+\alpha,
\qquad
(h\mu)P+\alpha=h\mu.
\]

The measure $\alpha$ is unique by uniqueness of the absolutely continuous
part of $\nu_sP$ relative to $\mu$. Consequently,
\[
\alpha=0
\quad\Longleftrightarrow\quad
\nu_sP=\nu_s
\quad\Longleftrightarrow\quad
(h\mu)P=h\mu.
\]
\end{proof}

The measure $\alpha$ records the one-sided transfer from the singular component
into the measure class of $\mu$.

The construction is relative to the chosen reference measure. If the roles of
$\mu$ and $\nu$ are reversed, the same argument yields the corresponding transfer
for the Lebesgue decomposition of $\mu$ with respect to $\nu$. The transfer therefore depends on the choice of the reference measure.

\subsection{Vanishing under finiteness}

The first mechanism forcing the transfer to vanish is conservation of
finite mass.

\begin{proposition}[Finite components have zero transfer]
\label{prop:finite-transfer}
Under the assumptions of Theorem~\ref{thm:transfer}, if either
\[
        (h\mu)(E)<\infty
        \qquad\text{or}\qquad
        \nu_s(E)<\infty,
\]
then
\[
        \alpha=0.
\]
Hence $h\mu$ and $\nu_s$ are separately invariant. If moreover
\[
        0<(h\mu)(E)<\infty,
\]
then
\[
        \frac{h\mu}{(h\mu)(E)}
\]
is an invariant probability measure.
\end{proposition}

\begin{proof}
If $(h\mu)(E)<\infty$, then evaluating
\[
(h\mu)P+\alpha=h\mu
\]
on $E$ gives
\[
(h\mu)P(E)+\alpha(E)=(h\mu)(E).
\]
Since $P$ preserves the total mass of the finite measure $h\mu$,
\[
(h\mu)P(E)=(h\mu)(E),
\]
and hence $\alpha(E)=0$. Thus $\alpha=0$.

The case $\nu_s(E)<\infty$ is identical: from
\[
\nu_sP=\nu_s+\alpha
\]
and preservation of the total mass of the finite measure $\nu_s$, we get
\[
\nu_s(E)+\alpha(E)=\nu_s(E),
\]
so again $\alpha=0$.

The remaining assertions are immediate.
\end{proof}

Thus
\[
        \alpha\neq0
        \quad\Longrightarrow\quad
        (h\mu)(E)=\nu_s(E)=\infty.
\]
The conclusion remains valid for  infinite components when the reference
measure is conservative.

\subsection{Conservativity kills transfer}
\label{subsec:conservative-transfer}

Recall that a sigma-finite invariant measure $\mu$ is
\emph{conservative} for $P$ if, for every $B\in\mathcal E$ with
$0<\mu(B)<\infty$,
\[
        P_x(\tau_B^+<\infty)=1
        \quad\text{for $\mu$-almost every }x\in B,
        \qquad
        \tau_B^+:=\inf\{n\geq1:X_n\in B\}.
\]
This is a standard recurrence formulation of conservativity for Markov
chains with a sigma-finite invariant measure; see, e.g.,
Revuz~\cite{Revuz} and Foguel~\cite{Foguel1969}.

\begin{lemma}[Conservativity implies potential divergence]
\label{lem:conservative-potential-divergence}
Let $\mu$ be a sigma-finite invariant measure which is conservative for
$P$. Then, for every $B\in\mathcal E$ with $0<\mu(B)<\infty$,
\[
\sum_{n=0}^{\infty}P^n(x,B)=\infty
\quad\text{for $\mu$-almost every }x\in B.
\]
\end{lemma}

\begin{proof}
Fix $B\in\mathcal E$ with $0<\mu(B)<\infty$, and set
\[
q(y):=P_y(\tau_B^+<\infty),
\qquad
G_B:=\{y\in B:q(y)=1\}.
\]
Conservativity gives $\mu(B\setminus G_B)=0$. Since $\mu P^n=\mu$,
\[
\int_E P^n(x,B\setminus G_B)\,\mu(dx)=0
\qquad (n\geq0),
\]
so
\[
A:=\bigcap_{n\geq0}\{x:P^n(x,B\setminus G_B)=0\}
\]
has full $\mu$-measure. Hence $A\cap G_B$ has full $\mu$-measure in $B$.

Fix $x\in A\cap G_B$. Then
\[
P_x(X_n\in B\setminus G_B)=0,
\qquad n\ge0.
\]
Let $\tau_B^{(1)}:=\tau_B^+$ and, for $k\ge1$,
\[
\tau_B^{(k+1)}
:=
\inf\{n>\tau_B^{(k)}:X_n\in B\}.
\]
The first return is finite almost surely because $x\in G_B$.

Assume that $\tau_B^{(k)}<\infty$ almost surely. Since the chain cannot
visit $B\setminus G_B$ at any deterministic time with positive
probability,
\[
P_x\!\left(
\tau_B^{(k)}<\infty,\,
X_{\tau_B^{(k)}}\in B\setminus G_B
\right)=0.
\]
Hence
\[
X_{\tau_B^{(k)}}\in G_B
\qquad P_x\text{-a.s. on }\{\tau_B^{(k)}<\infty\}.
\]
By the strong Markov property at $\tau_B^{(k)}$,
\[
P_x\!\left(
\tau_B^{(k+1)}<\infty
\mid
\mathcal F_{\tau_B^{(k)}}
\right)
=
q\!\left(X_{\tau_B^{(k)}}\right)
=
1
\qquad P_x\text{-a.s.}
\]
Thus every successive return time is finite almost surely.
Hence $X_n\in B$ infinitely often $P_x$-almost surely, and therefore
\[
\sum_{n=0}^{\infty}\mathbf 1_B(X_n)=\infty
\qquad P_x\text{-almost surely}.
\]
By monotone convergence,
\[
\sum_{n=0}^{\infty}P^n(x,B)
=
\mathbb E_x\!\left[
\sum_{n=0}^{\infty}\mathbf 1_B(X_n)
\right]
=
\infty.
\]
Since $A\cap G_B$ has full $\mu$-measure in $B$, the result follows.
\end{proof}

\begin{proposition}[Conservative reference measures have zero transfer]
\label{prop:conservative-transfer}
Under the assumptions of Theorem~\ref{thm:transfer}, if $\mu$ is
conservative for $P$, then $\alpha=0$. Hence $h\mu$ and $\nu_s$ are
separately invariant.
\end{proposition}

\begin{proof}
Suppose that $\alpha\neq0$. Iterating
\[
(h\mu)P+\alpha=h\mu
\]
gives, for every $n\geq1$,
\[
(h\mu)P^n+\sum_{k=0}^{n-1}\alpha P^k=h\mu,
\]
and hence
\[
\sum_{k=0}^{n-1}\alpha P^k\leq h\mu.
\]

Since $\mu$ is sigma-finite and $\alpha\ll\mu$, there exists
$B\in\mathcal E$ such that
\[
\alpha(B)>0,
\qquad
0<\mu(B)<\infty,
\qquad
(h\mu)(B)<\infty.
\]
By Lemma~\ref{lem:conservative-potential-divergence},
\[
\sum_{k=0}^{\infty}P^k(x,B)=\infty
\]
for $\mu$-almost every $x\in B$. Since $\alpha\ll\mu$ and
$\alpha(B)>0$, Tonelli's theorem yields
\[
\sum_{k=0}^{\infty}\alpha P^k(B)
=
\int_E \sum_{k=0}^{\infty}P^k(x,B)\,\alpha(dx)
\geq
\int_B \sum_{k=0}^{\infty}P^k(x,B)\,\alpha(dx)
=
\infty.
\]

On the other hand, evaluating
\[
(h\mu)P^n+\sum_{k=0}^{n-1}\alpha P^k=h\mu
\]
on $B$ gives
\[
\sum_{k=0}^{n-1}\alpha P^k(B)\leq(h\mu)(B).
\]
Letting $n\to\infty$,
\[
\sum_{k=0}^{\infty}\alpha P^k(B)\leq(h\mu)(B)<\infty,
\]
a contradiction. Hence $\alpha=0$.
\end{proof}

Thus conservativity preserves the relative Lebesgue decomposition of every
sigma-finite invariant measure:
\[
\mu\ \text{conservative}
\quad\Longrightarrow\quad
(h\mu)P=h\mu,
\qquad
\nu_sP=\nu_s.
\]
In particular, nonzero transfer requires a nonconservative reference measure.

\subsection{Nonzero transfer in the nonconservative regime}
\label{subsec:strip-transfer}

Nonzero transfer can occur when conservativity fails.

\begin{example}[Nonzero transfer for a random walk on a strip]
\label{ex:strip-transfer}
Consider
\[
E=\mathbb Z\times\{0,1\},
\qquad
a_j:=(j,0),
\qquad
s_j:=(j,1).
\]
Fix $p\in(1/2,1)$, set $q:=1-p$, and let
\[
\rho:=\sqrt{\frac pq},
\qquad
\gamma:=2\sqrt{pq}\in(0,1).
\]
For background on random walks on strips, see
Bolthausen and Goldsheid~\cite{BolthausenGoldsheid2000} and
Roitershtein~\cite{Roitershtein2008}.

On the lower layer set
\[
P(a_j,a_{j+1})=p,
\qquad
P(a_j,a_{j-1})=q,
\]
and on the upper layer
\[
P(s_j,s_{j+1})=P(s_j,s_{j-1})=\sqrt{pq},
\qquad
P(s_j,a_j)=1-\gamma.
\]
The lower layer
\[
A:=\mathbb Z\times\{0\}
\]
is absorbing, and its counting measure
\[
\mu:=\sum_{j\in\mathbb Z}\delta_{a_j}
\]
is invariant. Since $p>q$, the walk on $A$ is transient, so $\mu$ is not
conservative.

Define
\[
h\mu:=\sum_{j\in\mathbb Z}\rho^j\delta_{a_j},
\qquad
\nu_s:=\sum_{j\in\mathbb Z}\rho^j\delta_{s_j},
\qquad
\nu:=h\mu+\nu_s.
\]
Then $\nu_s\perp\mu$. Since
\[
\frac p\rho=q\rho=\sqrt{pq},
\qquad
\sqrt{pq}\,\rho=p,
\qquad
\sqrt{pq}\,\rho^{-1}=q,
\]
we have, for every $j\in\mathbb Z$,
\[
(h\mu)P(\{a_j\})
=
p\rho^{j-1}+q\rho^{j+1}
=
\gamma\rho^j,
\]
and hence
\[
(h\mu)P=\gamma h\mu.
\]
Similarly,
\[
\nu_sP(\{s_j\})
=
\sqrt{pq}\,\rho^{j-1}
+
\sqrt{pq}\,\rho^{j+1}
=
\rho^j,
\]
while the jump from $s_j$ to $a_j$ contributes
\[
(1-\gamma)\rho^j.
\]
Thus
\[
\nu_sP=\nu_s+(1-\gamma)h\mu.
\]
It follows that
\[
\nu P
=
\gamma h\mu+\nu_s+(1-\gamma)h\mu
=
\nu.
\]

The transfer measure relative to $\mu$ is therefore
\[
\alpha=(1-\gamma)h\mu
=
\bigl(1-2\sqrt{pq}\bigr)h\mu
\neq0.
\]
Thus the strip gives a nonconservative invariant reference measure with
nonzero transfer.
\end{example}

\section{Absorbing transfer and accessibility}
\label{sec:absorbing-accessibility}

Let A be absorbing and decompose $\mu$ as
$\mu=\mathbf 1_A\mu+\mathbf 1_{A^c}\mu$.

\subsection{Transfer across an absorbing boundary}

Recall that a measurable set $A\in\mathcal E$ is \emph{absorbing} if
\[
        P(x,A)=1,
        \qquad x\in A.
\]
Let $\mu$ be a sigma-finite invariant measure. The decomposition
\[
        \mu=1_A\mu+1_{A^c}\mu
\]
need not consist of invariant measures: although no mass can leave
$A$, mass may enter $A$ from $A^c$.

For $C\in\mathcal E$, define
\[
        \alpha_A(C)
        :=
        \int_{A^c}P(x,C\cap A)\,\mu(dx).
\]

\begin{proposition}[Transfer across an absorbing boundary]
\label{prop:absorbing-transfer}
Let $\mu$ be a sigma-finite invariant measure and let $A$ be absorbing.
Then
\[
        (1_A\mu)P+\alpha_A
        =
        1_A\mu,
\]
and
\[
        (1_{A^c}\mu)P
        =
        1_{A^c}\mu+\alpha_A.
\]
Moreover,
\[
        \alpha_A=0
        \quad\Longleftrightarrow\quad
        P1_A=1_A
        \qquad \mu\text{-a.e.}
\]
\end{proposition}

\begin{proof}
Since $A$ is absorbing,
\[
        P(x,C)=P(x,C\cap A),
        \qquad x\in A.
\]
Using invariance of $\mu$,
\[
\begin{aligned}
        (1_A\mu)P(C)+\alpha_A(C)
        &=
        \int_A P(x,C\cap A)\,\mu(dx)
        +
        \int_{A^c}P(x,C\cap A)\,\mu(dx)\\
        &=
        \int_E P(x,C\cap A)\,\mu(dx)\\
        &=
        \mu(C\cap A)\\
        &=
        1_A\mu(C).
\end{aligned}
\]
Thus
\[
        (1_A\mu)P+\alpha_A
        =
        1_A\mu.
\]
Since
\[
        \mu
        =
        1_A\mu+1_{A^c}\mu
\]
and $\mu P=\mu$, it follows that
\[
        (1_{A^c}\mu)P
        =
        1_{A^c}\mu+\alpha_A.
\]

Finally, since $A$ is absorbing,
\[
        P1_A=1
        \qquad\text{on }A.
\]
Thus
\[
        P1_A=1_A
        \quad\mu\text{-a.e.}
\]
is equivalent to
\[
        P(x,A)=0
        \quad\text{for $\mu$-almost every }x\in A^c.
\]
By the definition of $\alpha_A$, this is equivalent to
\[
        \alpha_A=0.
\]
\end{proof}

Call $P$  \emph{absorbing-harmonic with respect to $\mu$} if every
absorbing set $A$ satisfies
\[
        P1_A=1_A
        \qquad \mu\text{-a.e.}
\]
Equivalently, every absorbing boundary has zero transfer:
\[
        \alpha_A=0.
\]

This property is automatic for finite invariant measures. Indeed, if
$\mu(E)<\infty$ and $A$ is absorbing, then
\[
        P1_A\geq1_A,
\]
while invariance gives
\[
        \int_E P1_A\,d\mu
        =
        \int_E1_A\,d\mu.
\]
Hence
\[
        P1_A=1_A
        \qquad\mu\text{-a.e.}
\]

More generally, conservativity also rules out transfer across an
absorbing boundary.

\begin{proposition}[Conservativity prevents absorbing transfer]
\label{prop:conservative-absorbing}
Let $\mu$ be a sigma-finite invariant measure. If $P$ is conservative
with respect to $\mu$, then $P$ is absorbing-harmonic with respect to
$\mu$. Equivalently, for every absorbing $A$,
\[
        \alpha_A=0.
\]
\end{proposition}

\begin{proof}
Let $A$ be absorbing. Suppose that $\alpha_A\neq0$. Then
\[
        \int_{A^c}P(x,A)\,\mu(dx)>0.
\]
Therefore there exists $k\geq1$ such that
\[
        C_k
        :=
        \left\{
        x\in A^c:
        P(x,A)\geq\frac1k
        \right\}
\]
has positive $\mu$-measure.

By sigma-finiteness, choose $B\subset C_k$ such that
\[
        0<\mu(B)<\infty.
\]
Since $B\subset A^c$ and $A$ is absorbing,
\[
        \{X_1\in A\}\subset\{\tau_B^+=\infty\}.
\]
Hence, for every $x\in B$,
\[
        P_x(\tau_B^+=\infty)
        \geq P_x(X_1\in A)
        =P(x,A)
        \geq \frac1k.
\]
This contradicts conservativity, which gives
\[
        P_x(\tau_B^+<\infty)=1
\]
for $\mu$-almost every $x\in B$.

Hence $\alpha_A=0$. Proposition~\ref{prop:absorbing-transfer} then gives
\[
        P1_A=1_A
        \qquad\mu\text{-a.e.}
\]
for every absorbing $A$.
\end{proof}

\subsection{Resolvent accessibility and zero-potential sets}
\label{subsec:zero-potential}

Fix $\theta\in(0,1)$ and define the discrete resolvent
\[
        R_\theta
        :=
        (1-\theta)
        \sum_{n=1}^{\infty}
        \theta^{n-1}P^n.
\] Our convention is a positive-time one: the resolvent starts at
$n=1$, so accessibility is tested through strictly positive times.
This is the convention used throughout the paper.

For $B\in\mathcal E$, let
\[
        Z_B
        :=
        \{x\in E:R_\theta(x,B)=0\}.
\]
Since all coefficients in the resolvent are strictly positive,
\[
        R_\theta(x,B)=0
        \quad\Longleftrightarrow\quad
        P^n(x,B)=0
        \quad\text{for every }n\geq1.
\]
In particular, $Z_B$ does not depend on the choice of
$\theta\in(0,1)$.

The zero-potential set $Z_B$ is absorbing.

\begin{lemma}[Zero-potential sets are absorbing]
\label{lem:zero-potential}
For every $B\in\mathcal E$, the set $Z_B$ is absorbing.
\end{lemma}

\begin{proof}
Let $x\in Z_B$. Then
\[
        P^n(x,B)=0
        \qquad
        \text{for every }n\geq1.
\]
For every $n\geq1$,
\[
        0
        =
        P^{n+1}(x,B)
        =
        \int_E P(x,dy)P^n(y,B).
\]
Since the integrand is nonnegative,
\[
        P^n(y,B)=0
        \qquad
        \text{for $P(x,\cdot)$-almost every }y.
\]
Intersecting these full $P(x,\cdot)$-measure sets over
$n\geq1$, we obtain
\[
        y\in Z_B
        \qquad
        \text{for $P(x,\cdot)$-almost every }y.
\]
Thus
\[
        P(x,Z_B)=1,
\]
and $Z_B$ is absorbing.
\end{proof}

The kernel $P$ is called \emph{indecomposable} if there do not exist
two disjoint nonempty absorbing measurable sets.

A nonzero sigma-finite measure $\psi$ is called an
\emph{irreducibility measure} for $P$ if
\[
        \psi(B)>0
        \quad\Longrightarrow\quad
        R_\theta(x,B)>0
        \qquad
        \text{for every }x\in E.
\]
Equivalently,
\[
        \psi(B)>0
        \quad\Longrightarrow\quad
        Z_B=\varnothing.
\]
Again, the definition is independent of $\theta$.

Indecomposability alone does not imply irreducibility, even for a very
simple deterministic kernel.

\begin{example}[Indecomposable but not irreducible]
\label{ex:shift}
Let
\[
        E=\mathbb Z
\]
and define
\[
        P(n,\cdot)=\delta_{n+1}.
\]
Every nonempty absorbing set is a right-invariant subset of $\mathbb Z$:
if $n\in A$, then $n+k\in A$ for every $k\geq0$. Hence any two nonempty
absorbing sets intersect, and $P$ is indecomposable.

On the other hand, $P$ admits no nonzero irreducibility measure. Indeed,
let $\psi$ be any nonzero measure. Since
\[
        \mathbb Z=\bigcup_{m\in\mathbb Z}(-\infty,m],
\]
there exists $m\in\mathbb Z$ such that
\[
        \psi((-\infty,m])>0.
\]
Set
\[
        B:=(-\infty,m].
\]
For every $x>m$,
\[
        P^n(x,B)=0
        \qquad
        \text{for every }n\geq1,
\]
so $x\in Z_B$. Thus $Z_B\neq\varnothing$, and $\psi$ cannot be an
irreducibility measure.

\end{example}

\section{Indecomposability, irreducibility, and absorbing separation}
\label{sec:indecomposability}
Absorbing harmonicity and zero-potential sets yield the
indecomposability--irreducibility principle.

\subsection{Indecomposability and irreducibility}

\begin{theorem}[Absorbing harmonicity and irreducibility]
\label{thm:absorbing-harmonicity-irreducibility}
Let $P$ admit a nonzero sigma-finite invariant measure $\mu$, and assume
that $P$ is absorbing-harmonic with respect to $\mu$. Then
\[
        P\text{ is indecomposable}
        \quad\Longleftrightarrow\quad
        P\text{ is $\mu$-irreducible}.
\]
In particular, this equivalence holds whenever $P$ is conservative
with respect to $\mu$, and hence whenever $\mu$ is an invariant
probability measure.
\end{theorem}

\begin{proof}
Assume first that $\mu$ is an irreducibility measure. Let $A_1,A_2$ be
disjoint nonempty absorbing sets.

If $\mu(A_1)>0$, then no point of $A_2$ can reach $A_1$, so
\[
P^n(x,A_1)=0,
\qquad x\in A_2,\ n\geq1.
\]
Hence $A_2\subset Z_{A_1}$, contradicting irreducibility. If
$\mu(A_1)=0$, then $\mu(A_1^c)>0$, while absorbance of $A_1$ gives
\[
P^n(x,A_1^c)=0,
\qquad x\in A_1,\ n\geq1.
\]
Thus $A_1\subset Z_{A_1^c}$, again a contradiction. Therefore $P$ is
indecomposable.

Conversely, suppose that $P$ is indecomposable but $\mu$ is not an
irreducibility measure. Choose $B\in\mathcal E$ with
\[
\mu(B)>0,
\qquad
Z_B\neq\varnothing,
\]
and set $A:=Z_B$. By Lemma~\ref{lem:zero-potential}, $A$ is nonempty and
absorbing. Since $A=Z_B$,
\[
P(x,B)=0,
\qquad x\in A,
\]
whereas absorbing harmonicity gives
\[
P(x,A)=0
\qquad\text{for $\mu$-almost every }x\in A^c.
\]
Using invariance of $\mu$,
\[
\mu(A\cap B)
=
\int_E P(x,A\cap B)\,\mu(dx)
=
0.
\]
Indeed, the integrand vanishes on $A$, and on $A^c$ it is bounded by
$P(x,A)$, which vanishes $\mu$-almost everywhere. Thus
\[
\mu(B\cap A^c)=\mu(B)>0,
\]
so $\mu(A^c)>0$.

Now
\[
P1_A=1_A
\qquad\mu\text{-a.e.}
\]
by absorbing harmonicity. Since $\mu$ is invariant, this identity
iterates to
\[
P^n1_A=1_A
\qquad\mu\text{-a.e.},
\qquad n\geq1.
\]
Hence
\[
P^n(x,A)=0,
\qquad n\geq1,
\]
for $\mu$-almost every $x\in A^c$, and therefore
\[
R_\theta(x,A)=0
\]
there. Since $\mu(A^c)>0$, the set $Z_A$ is nonempty.

The sets $A$ and $Z_A$ are disjoint: absorbance of $A$ gives
$R_\theta(x,A)=1$ on $A$, whereas $R_\theta(x,A)=0$ on $Z_A$. Both are
absorbing, the latter by Lemma~\ref{lem:zero-potential}, contradicting
indecomposability. Thus $\mu$ is an irreducibility measure.

The final assertion follows from
Proposition~\ref{prop:conservative-absorbing}.
\end{proof}

\subsection{Absorbing separation and multiplicity}

We say that a measurable set $B$ carries a measure $\lambda$ if
\[
        \lambda(E\setminus B)=0.
\]

Mutually singular invariant measures can be realized on disjoint absorbing carriers.

\begin{theorem}[Absorbing separation]
\label{thm:absorbing-separation}
Let $\lambda_1$ and $\lambda_2$ be two nonzero mutually singular
sigma-finite invariant measures for $P$. Then there exist disjoint
absorbing measurable sets $B_1,B_2$ such that
\[
        \lambda_1(E\setminus B_1)=0,
        \qquad
        \lambda_2(E\setminus B_2)=0.
\]
\end{theorem}

\begin{proof}
Choose $A\in\mathcal E$ such that
\[
\lambda_1(A^c)=0,
\qquad
\lambda_2(A)=0,
\]
and set
\[
B_1:=Z_{A^c},
\qquad
B_2:=Z_A.
\]
Invariance of $\lambda_1$ gives
\[
\int_E R_\theta(x,A^c)\,\lambda_1(dx)
=
\lambda_1(A^c)
=
0,
\]
so $B_1$ carries $\lambda_1$. The same argument shows that $B_2$ carries
$\lambda_2$.

By Lemma~\ref{lem:zero-potential}, both $B_1$ and $B_2$ are absorbing.
Since
\[
R_\theta(x,A)+R_\theta(x,A^c)=1,
\qquad x\in E,
\]
the sets $Z_A$ and $Z_{A^c}$ are disjoint. Hence $B_1\cap B_2=\varnothing$.
\end{proof}

The same separation is obtained in Attali~\cite{AttaliClosedAbsorbing} in a topological setting; here the separating sets are the zero-potential sets of the resolvent.

Chen and Tweedie~\cite{ChenTweedie} study orthogonality and absorbing
sets in connection with Doeblin decompositions and structural
conditions on families of absorbing sets. Their results are formulated
under countable-generation and separation assumptions on the measurable
space. Theorem~\ref{thm:absorbing-separation} instead isolates the
pairwise separation of two mutually singular sigma-finite invariant
measures through the sets $Z_A$ and $Z_{A^c}$.

For invariant probabilities, a mutually singular invariant pair can be
recovered directly from nonuniqueness through the Jordan decomposition,
without passing through an ergodic decomposition.

\begin{corollary}[Multiplicity, uniqueness, and ergodicity]
\label{cor:multiplicity-separation}
If $P$ admits two distinct invariant probability measures, then it
admits two disjoint nonempty absorbing sets. Consequently, an
indecomposable Markov kernel admits at most one invariant probability
measure; whenever such a measure exists, it is ergodic.
\end{corollary}

\begin{proof}
Let $\mu\neq\nu$ be invariant probability measures and set
\[
\sigma:=\mu-\nu.
\]
Since $\sigma P=\sigma$,
\[
|\sigma|=|\sigma P|\leq|\sigma|P.
\]
Both finite measures have the same total mass, and therefore
\[
|\sigma|P=|\sigma|.
\]
The Jordan components
\[
\sigma^\pm=\frac{|\sigma|\pm\sigma}{2}
\]
are thus invariant.

Moreover, $\sigma(E)=0$ and $\sigma\neq0$, so
\[
\sigma^+(E)=\sigma^-(E)>0.
\]
After normalization, $\sigma^+$ and $\sigma^-$ yield two mutually
singular invariant probability measures. By
Theorem~\ref{thm:absorbing-separation}, they admit disjoint nonempty
absorbing carriers. Hence $P$ cannot be indecomposable.

It follows that an indecomposable kernel admits at most one invariant
probability measure. If such a measure $\pi$ were not ergodic, there
would exist $A\in\mathcal E$ with
\[
P1_A=1_A
\qquad \pi\text{-a.e.},
\]
and
\[
0<\pi(A)<1.
\]
The identity $P1_A=1_A$ $\pi$-a.e. makes the restrictions of $\pi$ to
$A$ and $A^c$ invariant. After normalization, they would therefore give
two distinct invariant probability measures, contradicting uniqueness.
\end{proof}

Finite-measures arguments in a similar setting appear in
Le Van and Stachurski~\cite{LeVanStachurski2004}. There nondecomposability is defined relatively to a reference measure whereas the present notion is intrinsic and based on absorbing sets.

Whenever an invariant probability measure $\pi$ exists,
Theorem~\ref{thm:absorbing-harmonicity-irreducibility} shows that
$\pi$-irreducibility is equivalent to indecomposability. The preceding
corollary then implies that, in this case, $\pi$ is the unique invariant
probability measure and is ergodic.

More generally, let $\mu$ be a nonzero sigma-finite invariant measure
and suppose that $P$ is conservative with respect to $\mu$. Then
Theorem~\ref{thm:absorbing-harmonicity-irreducibility} identifies
indecomposability with $\mu$-irreducibility.

This includes the classical irreducible Harris recurrent regime, in
which the invariant sigma-finite measure is unique up to multiplication
by a positive constant; see, for instance, Meyn and
Tweedie~\cite{MeynTweedie} or Nummelin~\cite{Nummelin}. After
normalization, the finite-mass case corresponds to positive Harris
recurrence, while the infinite-mass case includes null recurrence.

\section{Zero transfer and potential structure}
\label{sec:zero-transfer}

Let $\alpha$ denote the transfer measure associated with
\[
\nu=h\mu+\nu_s.
\]
Recall from Theorem~1 that
\[
\nu_sP=\nu_s+\alpha,
\qquad
(h\mu)P+\alpha=h\mu.
\]

\begin{theorem}[Zero transfer and absorbing separation]
\label{thm:zero-transfer-separation}
Let $\mu$ and $\nu$ be sigma-finite invariant measures, with
$\mu\neq0$, and write
\[
        \nu=h\mu+\nu_s,
        \qquad
        \nu_s\perp\mu,
\]
with $\nu_s\neq0$. Let $\alpha$ be the associated transfer measure.
Then the following are equivalent:
\begin{enumerate}
\item
\[
        \alpha=0;
\]
\item $h\mu$ and $\nu_s$ are separately invariant;
\item there exist disjoint absorbing measurable sets $B_\mu,B_s$ such
that
\[
        \mu(E\setminus B_\mu)=0,
        \qquad
        \nu_s(E\setminus B_s)=0.
\]
\end{enumerate}
\end{theorem}

\begin{proof}
The equivalence between (1) and (2) follows directly from
Theorem~\ref{thm:transfer}.

Assume (1). Then $\nu_s$ is invariant. Since
        $\mu\perp\nu_s$,
Theorem~\ref{thm:absorbing-separation} yields disjoint absorbing
measurable sets $B_\mu,B_s$ carrying $\mu$ and $\nu_s$, respectively.
Thus (3) holds.

Conversely, assume (3). Since $B_s$ is absorbing and carries $\nu_s$,
we have $\nu_sP(E\setminus B_s)=0$. Together with
$\nu_sP=\nu_s+\alpha$ and $\nu_s(E\setminus B_s)=0$, this shows that
$\alpha(E\setminus B_s)=0$, so $\alpha$ is carried by $B_s$.

On the other hand, $B_\mu$ carries $\mu$ and is disjoint from $B_s$,
hence $\mu(B_s)=0$. Since $\alpha\ll\mu$, we also have
$\alpha(B_s)=0$. Therefore
\[
        \alpha=0.
\]
\end{proof}

 Thus, when $\nu_s\neq0$, zero transfer is equivalent to the existence of disjoint absorbing carriers for $\mu$ and $\nu_s$.

If $P$ is conservative with respect to $\mu$, Proposition~\ref{prop:conservative-transfer}
and Theorem~\ref{thm:zero-transfer-separation} yield disjoint absorbing measurable sets
carrying $\mu$ and $\nu_s$ whenever $\nu_s\neq0$.

\subsection{An ergodic dichotomy}

Recall that a nonzero sigma-finite invariant measure $\nu$ is
\emph{ergodic} if every measurable set $A$ satisfying
\[
        P1_A=1_A
        \qquad \nu\text{-a.e.}
\]
has either $\nu(A)=0$ or $\nu(A^c)=0$.

\begin{corollary}[Ergodic dichotomy]
\label{cor:ergodic-dichotomy}
Let $\mu$ and $\nu$ be sigma-finite invariant measures and write
\[
\nu=h\mu+\nu_s,
\qquad
\nu_s\perp\mu.
\]
Assume that $\nu$ is ergodic. If the associated transfer vanishes, then
\[
h=0 \quad \mu\text{-a.e.}
\qquad\text{or}\qquad
\nu_s=0.
\]
Equivalently,
\[
\nu\perp\mu
\qquad\text{or}\qquad
\nu\ll\mu.
\]
In particular, the same dichotomy holds whenever $P$ is conservative
with respect to $\mu$.
\end{corollary}

\begin{proof}
If $\alpha=0$, Theorem~\ref{thm:transfer} shows that $h\mu$ and
$\nu_s$ are separately invariant. If both were nonzero, then
\[
\nu=h\mu+\nu_s
\]
would be a decomposition of $\nu$ into two nonzero mutually singular
invariant measures, contradicting ergodicity of $\nu$. Hence either
\[
h=0 \quad \mu\text{-a.e.}
\qquad\text{or}\qquad
\nu_s=0.
\]
The equivalent formulation follows immediately.

If $P$ is conservative with respect to $\mu$, then
Proposition~\ref{prop:conservative-transfer} gives $\alpha=0$.
\end{proof}

No proportionality between $\nu$ and $\mu$ follows without additional
assumptions on the reference measure.

\subsection{Cumulative transfer and the Riesz decomposition}

Iterating the transfer identities gives the following.

\begin{proposition}[Cumulative transfer]
\label{prop:cumulative-transfer}
Under the assumptions of Theorem~1, for every $n\geq1$,
\[
(h\mu)P^n+\sum_{k=0}^{n-1}\alpha P^k=h\mu,
\]
and
\[
\nu_sP^n
=
\nu_s+\sum_{k=0}^{n-1}\alpha P^k.
\]
Consequently,
\[
\sum_{k=0}^{n-1}\alpha P^k\leq h\mu.
\]
\end{proposition}

\begin{proof}
Iterating
\[
(h\mu)P+\alpha=h\mu,
\qquad
\nu_sP=\nu_s+\alpha,
\]
gives the two identities. The inequality is immediate from
$(h\mu)P^n\geq0$.
\end{proof}

The partial sums increase to
\[
G\alpha:=\sum_{k=0}^{\infty}\alpha P^k\leq h\mu.
\]
Since $h\mu$ is sigma-finite, there exists a unique nonnegative measure
$\rho$ such that
\[
h\mu=G\alpha+\rho.
\]
Indeed, since $G\alpha\leq h\mu$, there exists a measurable function
$g\in[0,1]$ such that $G\alpha=g\,h\mu$. Hence
\[
\rho:=(1-g)h\mu
\]
is the unique nonnegative measure satisfying $h\mu=G\alpha+\rho$.
Moreover,
\[
G\alpha=\alpha+(G\alpha)P.
\]
Applying $P$ to the decomposition of $h\mu$ and using
\[
(h\mu)P+\alpha=h\mu
\]
gives
\[
h\mu=G\alpha+\rho P.
\]
Comparison with
\[
h\mu=G\alpha+\rho
\]
and cancellation of $G\alpha$ on a finite-measure exhaustion yield
\[
\rho P=\rho.
\]
Thus
\[
h\mu=G\alpha+\rho,
\qquad
\rho P=\rho.
\]

This is the Riesz decomposition of the excessive measure $h\mu$ with
defect $\alpha$; see Getoor and Glover~\cite{GetoorGlover1984},
Fitzsimmons and Getoor~\cite{FitzsimmonsGetoor1992},
Getoor~\cite{Getoor1990}, and
Fitzsimmons and Getoor~\cite{FitzsimmonsGetoor2009}.

\section{Continuous time}
\label{sec:continuous-time}

Let $(P_t)_{t\ge0}$ be a measurable Markov semigroup, with
$(t,x)\mapsto P_t(x,B)$ measurable for every $B\in\mathcal E$.

\subsection{Resolvent accessibility}

Fix $\lambda>0$ and define the normalized resolvent kernel
\[
        R_\lambda(x,B)
        :=
        \int_0^\infty
        \lambda e^{-\lambda t}P_t(x,B)\,dt.
\]
For $B\in\mathcal E$, set
\[
        Z_B
        :=
        \{x\in E:R_\lambda(x,B)=0\}.
\]
Since the density $\lambda e^{-\lambda t}$ is strictly positive on
$(0,\infty)$,
\[
        R_\lambda(x,B)=0
        \quad\Longleftrightarrow\quad
        P_t(x,B)=0
        \quad\text{for Lebesgue-almost every }t>0.
\]
Consequently, $Z_B$ does not depend on the choice of $\lambda>0$.

We call the semigroup $\mu$-\emph{irreducible} if
\[
        \mu(B)>0
        \quad\Longrightarrow\quad
        Z_B=\varnothing,
\]
or equivalently,
\[
        \mu(B)>0
        \quad\Longrightarrow\quad
        R_\lambda(x,B)>0
        \qquad
        \text{for every }x\in E.
\]
We call the semigroup \emph{indecomposable} if there are no two
disjoint nonempty measurable sets which are absorbing for every
$P_t$, $t\geq0$.

\begin{lemma}[Resolvent zero sets are semigroup-absorbing]
\label{lem:ct-zero-absorbing}
For every $B\in\mathcal E$, the set $Z_B$ is absorbing for the whole
semigroup.
\end{lemma}

\begin{proof}
Let $x\in Z_B$. For every $s\geq0$,
\[
\begin{aligned}
        \int_E R_\lambda(y,B)\,P_s(x,dy)
        =
        \int_0^\infty
        \lambda e^{-\lambda t}P_{s+t}(x,B)\,dt 
        =
        e^{\lambda s}
        \int_s^\infty
        \lambda e^{-\lambda u}P_u(x,B)\,du.
\end{aligned}
\]
Since $R_\lambda(x,B)=0$ and the integrand defining
$R_\lambda(x,B)$ is nonnegative, the last integral vanishes. Hence
\[
        P_s(x,Z_B)=1,
        \qquad s\geq0.
\]
\end{proof}

Resolvent and potential methods for recurrent irreducible Markov chains
and processes are classical; see, in particular, Duflo~\cite{Duflo1970}.
For
every $n\geq1$, the semigroup property gives
\[
        R_\lambda^n(x,B)
        =
        \int_0^\infty
        \frac{\lambda^n t^{n-1}e^{-\lambda t}}
             {(n-1)!}
        P_t(x,B)\,dt.
\]
Both the exponential density defining $R_\lambda$ and the Gamma
density defining $R_\lambda^n$ are strictly positive on $(0,\infty)$.
Hence
\[
        R_\lambda^n(x,B)>0
        \quad\Longleftrightarrow\quad
        R_\lambda(x,B)>0.
\]
Therefore
\[
        Z_B
        =
        \{x:R_\lambda^n(x,B)=0
        \text{ for every }n\geq1\},
\]
and the semigroup is $\mu$-irreducible if and only if $R_\lambda$ is
$\mu$-irreducible.

Indecomposability is also unchanged by passage to the normalized
resolvent. Indeed, every set absorbing for the whole semigroup is
absorbing for $R_\lambda$. Conversely, suppose that $A_1$ and $A_2$
are disjoint nonempty absorbing sets for $R_\lambda$. Then
\[
        A_i\subset Z_{A_i^c},
        \qquad i=1,2.
\]
By Lemma~\ref{lem:ct-zero-absorbing}, each $Z_{A_i^c}$ is absorbing
for every $P_t$. Moreover, $Z_{A_1^c}$ and $Z_{A_2^c}$ are disjoint:
otherwise a point in their intersection would satisfy
\[
        R_\lambda(x,A_1^c)=R_\lambda(x,A_2^c)=0,
\]
which is impossible since $A_1^c\cup A_2^c=E$. Thus
\[
        R_\lambda\text{ is indecomposable}
        \quad\Longleftrightarrow\quad
        (P_t)_{t\geq0}\text{ is indecomposable}.
\]

If $\mu$ is invariant for the semigroup, then
\[
        \mu R_\lambda
        =
        \int_0^\infty
        \lambda e^{-\lambda t}\mu P_t\,dt
        =
        \mu.
\]
We may therefore apply the discrete
indecomposability--irreducibility principle directly to $R_\lambda$.

\begin{corollary}[Continuous-time conservative principle]
\label{cor:ct-conservative-principle}
Let $(P_t)_{t\geq0}$ admit a nonzero sigma-finite invariant measure
$\mu$. Fix $\lambda>0$ and assume that $R_\lambda$ is conservative
with respect to $\mu$. Then
\[
        (P_t)_{t\geq0}\text{ is indecomposable}
        \quad\Longleftrightarrow\quad
        (P_t)_{t\geq0}\text{ is $\mu$-irreducible}.
\]
\end{corollary}

\begin{proof}
The measure $\mu$ is invariant for $R_\lambda$. By the observations
above, $R_\lambda$ is indecomposable exactly when the semigroup is
indecomposable, and $R_\lambda$ is $\mu$-irreducible exactly when the
semigroup is $\mu$-irreducible. The conclusion follows from
Theorem~\ref{thm:absorbing-harmonicity-irreducibility} applied to
$R_\lambda$.
\end{proof}

\subsection{The transfer cocycle}

Let $\mu$ and $\nu$ be sigma-finite measures invariant for the whole
semigroup and write
\[
        \nu=h\mu+\nu_s,
        \qquad
        \nu_s\perp\mu.
\]
For each $t\geq0$, Theorem~\ref{thm:transfer} applied to the Markov kernel $P_t$ gives a unique
nonnegative measure $\alpha_t\ll\mu$ such that
\[
\nu_sP_t=\nu_s+\alpha_t,
\qquad
(h\mu)P_t+\alpha_t=h\mu.
\]
Thus all transfer identities remain within the cone of nonnegative measures;
in particular, no subtraction of possibly infinite measures is involved.

At $t=0$, the identity $P_0=I$ gives
\[
        \nu_s=\nu_s+\alpha_0.
\]
Since $\nu_s\perp\mu$ and $\alpha_0\ll\mu$, uniqueness of the
Lebesgue decomposition relative to $\mu$ yields
\[
        \alpha_0=0.
\]

The semigroup property imposes a cocycle structure on this family.

\begin{proposition}[Transfer cocycle]
For every $s,t\ge0$,
\[
        \alpha_{s+t}=\alpha_sP_t+\alpha_t.
\]
\end{proposition}

\begin{proof}
Since $\mu P_t=\mu$, absolute continuity with respect to $\mu$ is
preserved by $P_t$. Hence
\[
        \alpha_sP_t\ll\mu,
\]
and therefore
\[
        \alpha_t+\alpha_sP_t\ll\mu.
\]
Using the semigroup property,
\[
\begin{aligned}
        \nu_sP_{s+t}
        =(\nu_sP_s)P_t
        =(\nu_s+\alpha_s)P_t
        =\nu_s+\alpha_t+\alpha_sP_t.
\end{aligned}
\]
On the other hand,
\[
        \nu_sP_{s+t}=\nu_s+\alpha_{s+t}.
\]
Both expressions are Lebesgue decompositions of the same measure
relative to $\mu$, with singular part $\nu_s$ and absolutely
continuous parts $\alpha_{s+t}$ and
$\alpha_t+\alpha_sP_t$, respectively. Uniqueness of the Lebesgue
decomposition therefore gives
\[
        \alpha_{s+t}
        =
        \alpha_t+\alpha_sP_t.
\]
\end{proof}

\begin{lemma}[Measurability of the transfer family]\label{lem:mesurability}
For every $C\in\mathcal E$, the map
\[
        t\longmapsto \alpha_t(C),
        \qquad t\ge0,
\]
is measurable.
\end{lemma}

\begin{proof}
Since $\nu_s$ is sigma-finite, choose measurable sets
$F_m\uparrow E$ such that
\[
        \nu_s(F_m)<\infty
        \qquad\text{for every }m.
\]
Fix $C\in\mathcal E$ and set $C_m=C\cap F_m$.
By the joint measurability assumption on the semigroup and Tonelli's
theorem,
\[
        t\longmapsto
        (\nu_sP_t)(C_m)
        =
        \int_E P_t(x,C_m)\,\nu_s(dx)
\]
is measurable. Since $\nu_s(C_m)<\infty$ and
\[
        (\nu_sP_t)(C_m)
        =
        \nu_s(C_m)+\alpha_t(C_m),
\]
it follows that
\[
        t\longmapsto\alpha_t(C_m)
\]
is measurable.

For each fixed $t$, continuity from below gives
\[
        \alpha_t(C_m)\uparrow\alpha_t(C).
\]
Hence $t\mapsto\alpha_t(C)$ is measurable as the pointwise limit of
measurable functions.
\end{proof}

\subsection{Conservativity kills the transfer cocycle}

The discrete conservative argument is applied to the
normalized resolvent rather than to an individual $P_t$, which need
not be conservative.

\begin{theorem}[Conservative resolvents have zero transfer]
\label{thm:ct-conservative-transfer}
Let $\mu$ and $\nu$ be sigma-finite measures invariant for
$(P_t)_{t\geq0}$, with $\mu\neq0$, and write
\[
        \nu=h\mu+\nu_s,
        \qquad
        \nu_s\perp\mu.
\]
Fix $\lambda>0$. If the normalized resolvent $R_\lambda$ is
conservative with respect to $\mu$, then
\[
        \alpha_t=0
        \qquad
        \text{for every }t\geq0.
\]
Consequently, $h\mu$ and $\nu_s$ are separately invariant for the
whole semigroup.
\end{theorem}

\begin{proof}
The measures $\mu$ and $\nu$ are invariant for $R_\lambda$, and their
relative Lebesgue decomposition remains
\[
\nu=h\mu+\nu_s.
\]
Conservativity of $R_\lambda$ with respect to $\mu$ and
Proposition~\ref{prop:conservative-transfer} therefore give
\[
\nu_sR_\lambda=\nu_s.
\]

For $C\in\mathcal E$, set
\[
\beta(C):=
\int_0^\infty
\lambda e^{-\lambda t}\alpha_t(C)\,dt.
\]
Lemma~\ref{lem:mesurability} and Tonelli's theorem show that $\beta$ is a
nonnegative measure satisfying
\[
(\nu_sR_\lambda)(C)=\nu_s(C)+\beta(C).
\]

Let $E_m\uparrow E$ with $\nu_s(E_m)<\infty$. On each $E_m$,
\[
\nu_s(E_m)
=
(\nu_sR_\lambda)(E_m)
=
\nu_s(E_m)+\beta(E_m),
\]
and hence $\beta(E_m)=0$. Continuity from below yields $\beta=0$, so
\[
\int_0^\infty
\lambda e^{-\lambda t}\alpha_t(E)\,dt=0.
\]
As the integrand is nonnegative,
\[
\alpha_t(E)=0
\]
for Lebesgue-almost every $t>0$. Since each $\alpha_t$ is a nonnegative
measure, this is equivalent to $\alpha_t=0$ for Lebesgue-almost every
$t>0$. Together with $\alpha_0=0$, this gives almost-everywhere
vanishing on $[0,\infty)$.

It remains to rule out exceptional times. If $\alpha_{t_0}\neq0$ for
some $t_0\ge0$, the cocycle identity gives
\[
\alpha_{s+t_0}
=
\alpha_sP_{t_0}+\alpha_{t_0}
\geq
\alpha_{t_0},
\qquad s\ge0.
\]
Thus the transfer would be nonzero for every $t\ge t_0$, contradicting
its almost-everywhere vanishing. Hence
\[
\alpha_t=0,
\qquad t\ge0.
\]

The identities
\[
\nu_sP_t=\nu_s+\alpha_t,
\qquad
(h\mu)P_t+\alpha_t=h\mu
\]
then give separate invariance of $\nu_s$ and $h\mu$ for the whole
semigroup.
\end{proof}

\begin{remark}[Conservativity is sufficient, not necessary]
Conservativity is not the only sufficient condition for zero transfer.
Zero transfer may also follow from smoothing: if, for
some $t_0>0$,
\[
        \eta P_{t_0}\ll\mu
\]
for every sigma-finite measure $\eta$, then every sigma-finite invariant
measure $\nu$ satisfies
\[
        \nu=\nu P_{t_0}\ll\mu.
\]
Hence its singular component relative to $\mu$ vanishes, and so the
associated transfer is identically zero.

For the Brownian semigroup on $\mathbb R^d$,
Lebesgue measure $dx$ is sigma-finite and invariant, and for every $t>0$
the transition kernel admits a strictly positive Gaussian density with
respect to $dx$. Hence, for every sigma-finite invariant measure $\nu$,
\[
        \nu=\nu P_t\ll dx.
\]
Thus the singular component of $\nu$ relative to $dx$ vanishes, and so
the associated transfer is identically zero in every dimension.

For $d=1,2$, Brownian motion is recurrent and the normalized resolvent
is conservative with respect to $dx$, so Theorem~\ref{thm:ct-conservative-transfer} applies directly.
For $d\ge3$, Brownian motion is transient and the normalized resolvent
is not conservative, yet the transfer still vanishes because of the
smoothing property above.

Thus conservativity is sufficient but not necessary for zero
transfer; Example~\ref{ex:strip-transfer} shows that its failure can permit nonzero transfer. 

\end{remark}

\subsection{Common absorbing separation}

\begin{proposition}[Simultaneous absorbing separation]
\label{prop:ct-singular-separation}
Let $\lambda_1$ and $\lambda_2$ be two nonzero mutually singular
sigma-finite measures invariant for $(P_t)_{t\geq0}$. Then there exist
two disjoint measurable sets $B_1,B_2$, absorbing for the whole
semigroup, such that
\[
        \lambda_1(E\setminus B_1)=0,
        \qquad
        \lambda_2(E\setminus B_2)=0.
\]
\end{proposition}

\begin{proof}
Choose $A\in\mathcal E$ such that
\[
        \lambda_1(A^c)=0,
        \qquad
        \lambda_2(A)=0.
\]
Set
\[
        B_1:=Z_{A^c},
        \qquad
        B_2:=Z_A.
\]

Since $\lambda_1$ is invariant for the semigroup,
\[
        \lambda_1R_\lambda=\lambda_1.
\]
Hence
\[
        \int_E R_\lambda(x,A^c)\,\lambda_1(dx)
        =
        \lambda_1(A^c)
        =
        0,
\]
so $B_1$ carries $\lambda_1$. Similarly, $B_2$ carries $\lambda_2$.

By Lemma~\ref{lem:ct-zero-absorbing}, both sets are absorbing for the
whole semigroup. Finally,
\[
        R_\lambda(x,A)+R_\lambda(x,A^c)=1
\]
for every $x\in E$, so
\[
        B_1\cap B_2=\varnothing.
\]
\end{proof}

\begin{theorem}[Continuous-time zero transfer and absorbing separation]
\label{thm:ct-zero-transfer-separation}
Let $\mu$ and $\nu$ be sigma-finite measures invariant for
$(P_t)_{t\geq0}$, with $\mu\neq0$, and write
\[
        \nu=h\mu+\nu_s,
        \qquad
        \nu_s\perp\mu,
\]
with $\nu_s\neq0$. Let $(\alpha_t)_{t\geq0}$ be the associated transfer
family. Then the following are equivalent:
\begin{enumerate}
\item
\[
        \alpha_t=0
        \qquad
        \text{for every }t\geq0;
\]
\item $h\mu$ and $\nu_s$ are separately invariant for the whole
semigroup;
\item there exist disjoint measurable sets $B_\mu,B_s$, absorbing for
the whole semigroup, such that
\[
        \mu(E\setminus B_\mu)=0,
        \qquad
        \nu_s(E\setminus B_s)=0.
\]
\end{enumerate}
\end{theorem}

\begin{proof}
The equivalence between (1) and (2) follows from
\[
        \nu_sP_t=\nu_s+\alpha_t,
        \qquad
        (h\mu)P_t+\alpha_t=h\mu.
\]

Assume (1). Then $\nu_s$ is invariant for the whole semigroup. Since
\[
        \mu\perp\nu_s,
\]
Proposition~\ref{prop:ct-singular-separation} gives disjoint common
absorbing carriers $B_\mu$ and $B_s$. Thus (3) holds.

Conversely, assume (3). Since $B_s$ is absorbing and carries $\nu_s$,
\[
        \nu_sP_t(E\setminus B_s)=0
        \qquad
        \text{for every }t\geq0.
\]
Using
\[
        \nu_sP_t=\nu_s+\alpha_t
\]
and
\[
        \nu_s(E\setminus B_s)=0,
\]
we see that $\alpha_t$ is carried by $B_s$. Since
\[
        \alpha_t\ll\mu
        \qquad\text{and}\qquad
        \mu(B_s)=0,
\]
it follows that
\[
        \alpha_t=0
        \qquad
        \text{for every }t\geq0.
\]
\end{proof}

By Theorems~\ref{thm:ct-conservative-transfer} and
\ref{thm:ct-zero-transfer-separation}, if $R_\lambda$ is conservative
with respect to $\mu$ and $\nu_s\neq0$, then the reference and singular
components admit disjoint measurable carriers that are absorbing for
the whole semigroup.

\section{Conclusion}
\label{sec:conclusion}

The relative Lebesgue decomposition of two sigma-finite invariant
measures is not in general preserved by a Markov kernel. The singular component can transfer mass into the measure
class of the reference measure. This transfer vanishes when the
reference measure is conservative.

The same principle appears for absorbing sets. Conservativity prevents
mass from crossing an absorbing boundary and, through the zero-potential
sets of the resolvent, makes indecomposability equivalent to
irreducibility with respect to the invariant measure.

Zero transfer is also equivalent to separate invariance of the two
components and, when the singular component is nonzero, to their
separation on disjoint absorbing sets. In continuous time, the transfer
forms a positive cocycle, which vanishes whenever the normalized
resolvent is conservative.

Neither conservativity nor its failure alone determines whether transfer
occurs: the strip example has nonzero transfer, whereas Brownian
smoothing forces zero transfer even in the transient dimensions.








\bibliography{cdm-PA}

\end{document}